# Enhanced electron correlations in the new binary stannide PdSn$_4$: a homologue of the Dirac nodal arc semimetal PtSn$_4$

C. Q. Xu,[1,2] W. Zhou,[1] R. Sankar,[3] X. Z. Xing,[4] Z. X. Shi,[4] Z. D. Han,[1] B. Qian,[1**] J. H. Wang,[5] Zengwei Zhu,[5] J. L. Zhang,[6] A. F. Bangura[7], N. E. Hussey[8] and Xiaofeng Xu,[1,2]*

[1]*Advanced Functional Materials Lab and Department of Physics, Changshu Institute of Technology, Changshu 215500, China*

[2]*Department of Physics, Hangzhou Normal University, Hangzhou 310036, China*

[3]*Center for Condensed Matter Sciences, National Taiwan University, Taipei 10617, Taiwan*

[4]*School of Physics and Key Laboratory of MEMS of the Ministry of Education, Southeast University, Nanjing 211189, China*

[5]*Wuhan National High Magnetic Field Center, School of Physics, Huazhong University of Science and Technology, Wuhan, 430074, China*

[6]*High Magnetic Field Laboratory, Chinese Academy of Sciences, Hefei 230031, China*

[7]*Max-Planck-Institut fur Festkorperforschung, Heisenbergstr. 1, D-70569 Stuttgart, Germany*

[8]*High Field Magnet Laboratory (HFML-EMFL), Radboud University, Toernooiveld 7, 6525ED Nijmegen, Netherlands*

(Dated: June 16, 2017)

**The advent of nodal-line semi-metals, i.e. systems in which the conduction and valence bands cross each other along a closed trajectory (line or loop) inside the Brillouin zone, has opened up a new arena for the exploration of topological condensed matter in which, due to a vanishing density of states near the Fermi level, electron correlation effects may also play an important role. In spite of this conceptual richness however, material realization of nodal-line (loop) fermions is rare, with PbTaSe$_2$, ZrSiS and PtSn$_4$ the only promising known candidates. Here we report the synthesis and physical properties of a new compound PdSn$_4$ that is isostructural with PtSn$_4$ yet possesses quasiparticles with significantly enhanced effective masses. In addition, PdSn$_4$ displays an unusual polar angular magnetoresistance which at a certain field orientation, varies linearly with field up to 55 Tesla. Our study suggests that, in association with its homologue PtSn$_4$ whose low-lying excitations were recently claimed to possess Dirac node arcs, PdSn$_4$ may be a promising candidate in the search for novel topological states with enhanced correlation effects.**

I.     Introduction

The quest for quantum materials with diverse symmetry-protected topological states has been a major driver of the field of condensed matter physics in recent years [1,2]. Due to the advent of Dirac/Weyl semimetals, the research interest in topological phenomena now encompasses topological insulators (TIs) and semimetals [3,4]. Three-dimensional (3D) topological Dirac or Weyl semimetals are new states of matter that possess a linear band dispersion in the bulk along all three momentum directions, whose low-energy quasiparticles are the condensed matter analogues of Dirac and Weyl fermions in relativistic high energy physics. Many exotic transport behaviors associated with these topological states, such as the chiral anomaly and large linear magnetoresistance (LMR), were identified experimentally soon after their discovery [5-10].

More recently, the focus has shifted towards a new breed of topological materials, namely the topological nodal-line (or nodal loop) semimetals (NLSMs), in which the Dirac node extends along a curve (or loop) in momentum space, in contrast with the isolated Dirac point node present in TIs and Dirac semimetals [11-17]. Their interest arises primarily from the possibility that the topological aspects of NLSMs may be distinct from those associated with TIs or WSMs [2,18]. It is claimed, for example, that due to the vanishing density of states near the Fermi level $\varepsilon_F$, screening of the Coulomb interaction is weaker in NLSMs than in conventional metals and remain long-ranged [19]. This, coupled with their metallic nature, could make NLSMs more susceptible to various types of order (e.g. superconductivity, magnetism or charge order), both in the bulk [20] and on the surface [21].

In this article, we report the synthesis and physical characterization of a newly synthesized compound $PdSn_4$ that is isostructural with the Dirac nodal arc semimetal $PtSn_4$ [16,17]. Unlike a host of other topological materials that are predicted theoretically first and subsequently verified by experiment, $PtSn_4$ was synthesized many years before its classification as a topological material. Inspired by the observation of an ultrahigh magnetoresistance (MR) in high quality single crystals [16], detailed angle-resolved photoemission spectroscopy (ARPES) studies subsequently revealed a novel topological phase in $PtSn_4$: surface-derived Dirac nodes that formed three open lines ('arcs') in reciprocal space, one at the Fermi level $\varepsilon_F$ and the other two at a binding energy of 60 meV [17]. These Dirac nodal arcs are thought to be responsible for the observed ultrahigh MR. It is not yet clear, however, how robust this nodal arc structure in $PtSn_4$ is to variations in the strength of the spin-orbit coupling (SOC) or to the presence of strong electron correlations, and how fragile these topological nodal-line fermions are to other instabilities such as charge density waves (e.g. as in another NLSM ZrSiS [22]). According to band structure calculations, the Sn $p$ states in PtSn$_4$ account for most of the density of states at $\varepsilon_F$, with only a minor contribution from Pt-derived $d$ bands [16]. Therefore, replacing Pt with the lighter element Pd is not expected to change the band structure nor the Dirac nodal structure significantly.

With this in mind, we have synthesized the first single crystals of PdSn$_4$, a sister compound of PtSn$_4$ with a nominally reduced SOC, and carried out a comprehensive investigation of its bulk physical properties. A large linear MR $\Delta\rho/\rho$, reaching over ~ 1300% at 2 K under a 9 T field, has been observed along certain field orientations, contrasting markedly with the quadratic MR observed in PtSn$_4$ [16]. The angular dependence of magnetoresistance (AMR) is found to possess a strong anisotropic structure which is smeared out gradually with increasing $T$. The temperature under which the symmetry of the AMR undergoes salient change coincides with the characteristic temperatures where the Hall resistivity, thermoelectric power (TEP) and magnetization also show significant variations, implying either a modification of the Fermi surface with decreasing $T$ or a (partial) gapping of carriers.

In addition, clear de Haas-van Alphen oscillations are revealed in magnetization isotherms. While the main frequencies are found to be almost identical to those of PtSn$_4$, the effective masses are enhanced by a factor of 3-4. These heavier masses are reflected in the enhanced coefficient of the $T^2$ resistivity, the enhanced electronic specific heat coefficient $\gamma$ and the enhanced low-$T$ thermopower. Overall, these findings reveal the presence of sizeable electron correlations in PdSn$_4$ relative to PtSn$_4$, possibly due to coupling to some form of density wave. Based on these observations, we identify PdSn$_4$ as a candidate material, in addition to ZrSiS, for hosting novel topological states with strong correlations, an ingredient on which much less attention has been paid thus far in the nascent field of topological physics.

## II.     Results

As shown in Fig. S1 of the Supplementary Information (SI), all X-ray diffraction peaks obtained on our single crystals can be indexed with the same orthorhombic (Ccca, No. 68) space group structure as PtSn$_4$ [16]. The crystal structure itself is composed of stacks of Pd atoms sandwiched between two coplanar Sn layers that lie within the $ac$ crystallographic plane (see SI). Figure 1 shows the characteristic temperature dependence of the resistivity $\rho(T)$ of PdSn$_4$ under various magnetic fields **H**//$b$, (i.e. perpendicular to the basal plane). The room temperature resistivity has a value of ~ 20 $\mu\Omega$cm and the (zero-field) residual resistivity ratio (RRR = $\rho$(300 K)/$\rho$(2 K)) reaches values as high as 100, indicating a high degree of metallicity. From room temperature down to 50 K, $\rho(T)$ follows an approximately linear $T$-dependence.

Below 50 K, however, $\rho(T)$ crosses over to a quadratic $T$-dependence, i.e. $\rho(T) = \rho_0 + AT^2$ – behavior that is more characteristic of correlated metals than TIs. The coefficient $A$ is estimated as 7 x 10$^{-4}$ $\mu\Omega$cm/K$^2$, which is a factor of 3 higher than has been reported for PtSn$_4$ [16]. This high degree of metallicity is reflected in the size of the transverse MR which at $T$ = 2 K reaches about 660% under 9 T when the field is applied along the crystalline $b$-axis.

The specific heat $C$ of PdSn$_4$ is presented in Fig. 1(b). Overall, $C(T)$ of PdSn$_4$ is very similar to that of PtSn$_4$ [16]. Within the measured temperature range, no signature of a thermodynamic phase transition can be resolved. Through fitting the low-$T$ $C(T)$ data to the formula $C(T) = \gamma T + \beta_3 T^3 + \beta_5 T^5$, the Sommerfeld (electronic) coefficient $\gamma$ and the harmonic (phonon) term $\beta_3$ can be extracted. For PdSn$_4$, we find $\gamma$ = 7 mJ/mol.K$^2$, compared to 4 mJ/mol.K$^2$ for PtSn$_4$. This larger value for $\gamma$ is consistent with the enhanced $A$ coefficient in the low-$T$ resistivity, suggesting that electron correlation effects are indeed more prominent in PdSn$_4$ than in PtSn$_4$. In many correlated metals, the ratio $A/\gamma^2$ (known as the Kadowaki-Woods ratio or KWR) is found to take a nearly universal value of 10 $\mu\Omega$cm(mol.K/J)$^2$ [23]. For PdSn$_4$, KWR ~ 14 $\mu\Omega$cm(mol.K/J)$^2$. While this value lies close to the value typically found in strongly correlated electron systems such as the heavy fermions, it should be noted that the KWR itself is very sensitive to a host of material-specific factors such as carrier density, multi-band effects and its effective dimensionality [24,25]. Nonetheless, we note that the KWRs in PdSn$_4$ and PtSn$_4$ are comparable in magnitude, implying that any enhancement in $A$ and $\gamma$ is likely due to correlation effects rather than to differences in their Fermi surface topologies. Finally, from the coefficient of the $T^3$ term $\beta$ = 0.8 mJ/mol.K$^4$, we obtain the Debye temperature $\theta_D$ ~ 230 K, similar to the one obtained for PtSn$_4$ (210 K).

The Hall resistivity $\rho_{xy}(B)$ and thermoelectric power (TEP) of PdSn$_4$ single crystals are shown in Figures 2 and 3 respectively. At high temperatures, $\rho_{xy}$ is positive and increases linearly with $B$, implying dominant hole-like transport. Below $T$~ 25 K, however, the $\rho_{xy}(B)$ curve starts to develop non-linearity and, as shown in the inset, $\rho_{xy}(B)$ at the lowest temperatures, passes through two inflexion points. This double-inflexion-point structure of the Hall resistivity is a rare phenomenon, even in multiband systems. A similar Hall response has been seen, for example, in PdCrO$_2$ where it was ascribed to the anomalous Hall effect induced by antiferromagnetic correlations [26]. It is presumed that in non-magnetic PdSn$_4$, this peculiar Hall response will have a different origin, though further discussion on this point lies beyond the scope of the present study.

Figure 3(a) shows the *T*-dependent TEP over a broad temperature range. As is clear from the figure, magnetic fields appear to have a negligible effect on the TEP, in marked contrast with the large field-dependent TEP found in PtSn$_4$ [16]. This difference, however, may lie in the different carrier mobilities of the two homologues, as reflected, for example, in their respective RRRs. Intriguingly, throughout the whole temperature range studied, the TEP of PdSn$_4$ has an opposite sign to the main feature of the Hall resistivity. Such opposite signs between TEP and Hall resistivity have been previously seen in the candidate TI LaAgSb$_2$ [27] as well as in the ferromagnetic superconductor UCoGe [28]. Combined with the highly non-linear $\rho_{xy}(B)$, this observation suggests a rather complex band structure in PdSn$_4$ near the Fermi level. Upon cooling, the TEP of PdSn$_4$ shows a kink at $T \sim 25$ K, i.e. at the same temperature where the $\rho_{xy}(B)$ curves start to become non-linear. Perhaps more significantly, the slope of the low-*T* TEP, $S/T$ = -0.096 $\mu$V/K, is found to be approximately one order of magnitude higher than reported in PtSn$_4$ (where, incidentally, the TEP was found to be positive at the lowest temperatures). From this low-*T* slope, we obtain a corresponding '$q$' value - linking the TEP to the electronic specific heat via the expression $q = S/T \, (N_{av} e / \gamma)$ - of 1.4, that falls precisely within the range $0.5 < |q| < 2.0$ found for a host of strongly correlated electron systems [29]. Thus, both the KWR and '$q$' values for PdSn$_4$ appear to be consistent with expectations for correlated metals. We are certainly not aware of other semimetals which give comparable values.

The *T*-dependent susceptibility, measured under a field of 0.5 T, is shown in Fig. 3(b). Similar to PtSn$_4$, the sample is weakly diamagnetic. Around $T$ = 25 K, the susceptibility undergoes a rapid increase. Intriguingly, the whole shape of the *M*(*T*) curve exhibits a similar profile to that of TEP, suggesting that the same electrons, which participate in the entropy transport, are responsible for the magnetization. It is worth noting too that the characteristic temperatures (~ 25 K) identified from the abnormal changes in *T*- dependent $\rho_{xy}(B)$, TEP and susceptibility coincide the temperature at which the symmetry of the AMR, to be discussed in more detail below, also changes, signifying a possible change in the Fermi surface topology across this temperature. Unfortunately, comparison of the absolute magnitude of the Pauli susceptibility $\chi_P$ with PtSn$_4$ is not possible until the additional orbital and van Vleck contributions in both systems are determined independently and subtracted.

The magnetization curves also exhibit clear de Haas-van Alphen (dHvA) oscillations (see Fig. 4(a)) which can provide, in principle, a direct measure of the quasiparticle effective masses for the different bands in PdSn$_4$.

A fast Fourier transform (FFT) of the dHvA oscillations is shown in Fig. 4(b). Several independent frequencies are identified, corresponding to different extremal Fermi surface cross-sectional areas $A_e$ within the Brillouin zone. We label the three most prominent peaks in sequence as $\alpha$, $\beta$ and $\gamma$ with corresponding frequencies ($F$) of 50 T, 66 T and 125 T. These values are very similar to the main frequencies found in PtSn$_4$ [16], suggesting a comparable Fermi surface topology (at low temperatures). The corresponding areas $A_e$, and Fermi wave vectors $k_F$, obtained from the Onsager relation $F = (\hbar/2\pi e)A_e$, are listed in Table 1.

In order to obtain the effective mass $m^*$ for each orbit, the oscillatory dHvA data are fitted at different temperatures to the standard Lifshitz-Kosevich formula [30]

$$\Delta M \propto -R_T \cdot R_D \cdot \sin\left[2\pi\left(\frac{F}{B} + \frac{1}{2} - \frac{\phi_B}{2\pi} - \delta\right)\right]$$

where $R_T$ ($R_D$) is the thermal damping factor (Dingle damping factor), $\phi_B$ is the Berry phase and $\delta$ is an additional phase shift determined by the dimensionality of Fermi surface (see SI for the discussion of the Berry phase). Therein,

$$R_T = \frac{2\pi^2 k_B T m^*/eB\hbar}{\sinh(2\pi^2 k_B T m^*/eB\hbar)}$$

and $R_D = \exp(2\pi^2 k_B T_D m^*/eB\hbar)$, where $k_B$ is the Boltzmann constant. As shown in Figs. 4(c-e), the best fits for each frequency yield $m^*$ values of 0.33 $m_e$, 0.41 $m_e$ and 0.54 $m_e$ for the $\alpha$, $\beta$ and $\gamma$ pockets respectively, where $m_e$ is the bare electron mass. The corresponding values for $T_D$, the Fermi velocity $v_F$, the Fermi energy $\varepsilon_F$, the mean free path $\ell$ and the carrier mobility $\mu$ are also listed in Table 1 for completeness.

Consistent with the bulk physical properties, the obtained effective masses are found to be significantly larger than in PtSn$_4$ where for the $\alpha$, $\beta$ and $\gamma$ pockets (with respective frequencies of 52, 63 and 84 T), the $m^*$ values were all in the range 0.11 - 0.16 $m_e$ [16]. Since the mass obtained from dHvA oscillations is equivalent to the thermodynamic mass and proportional to the rate of change in Fermi surface area with energy, the fact that the masses are 3-4 times larger for pockets of the same area imply that the bands near $\varepsilon_F$ must be significantly narrower in PdSn$_4$ than in PtSn$_4$. Detailed analysis of the oscillatory signals in the SI also provides evidence for a possible non-trivial Berry phase, though verification of the full Dirac-like band structure in PdSn$_4$ will require detailed ARPES studies in due course.

Finally, we turn to discuss the anomalous angular magnetoresistance (AMR) of $PdSn_4$. The AMR measured on a second sample (with RRR ~ 87) under two distinct *B*-rotation configurations are represented in Fig. 5 and 6, respectively. For the first rotation configuration (referred to as *B//bI* and depicted schematically in Fig. 5(d)), the field is rotated from the *b*-axis to *B//I*. As seen from the polar plots of AMR in Fig. 5(a), the angle ($\theta$) dependence of the resistivity under a field of 9 T shows a distinct butterfly-like shape at low *T*, implying a strongly anisotropic Fermi-surface. The maximum MR is found at a tilted angle of $\theta$ ~ 20°. In comparison, at $\theta$ = 0° and 90°, MR is relatively small, resulting in the appearance of 'two dips'. With increasing temperature, these two dips fade away progressively and above *T* ~ 30 K, the dip at $\theta$ = 0° disappears altogether. The remaining dip at $\theta$ = 90° makes the polar plot of the AMR resemble a dumbbell with two-fold symmetry. In addition, the *B*-dependence of the MR for different $\theta$ values shows a variety of forms (see Fig. 5 (b)). When $\theta$ ~ 20°, i.e., the peak angle in the AMR, the MR grows linearly with *B* from almost zero field to the highest field studied (55 T), as shown in Fig. 5(c). On the other hand, when $\theta$ is aligned with the dip positions, the MR exhibits downward curvature at low fields while the overall shape remains quite linear. With increasing *T*, this downward curvature is gradually replaced by the more conventional quadratic low-*B* MR. The size of the MR at 9 T is also quickly suppressed with increasing *T*, consistent with the above $\rho(T)$ measurements under different fields in Fig. 1 (corresponding to $\theta$ = 0°).

For the other *B*-rotation configuration, which we denote as *B*⊥*I*, meaning that the field remains in the plane perpendicular to the current (see Fig. 6(d)), the AMR evolves with *T* in the same fashion (see Fig. 6(a)). One difference however is that the maximum MR occurs at angle $\theta$ close to 45°. The AMR symmetry at 2 K is almost four-fold. In addition, the critical temperature at which the dip at $\theta$ = 0° disappears shifts to ~ 40 - 50 K, which is slightly higher than the first configuration. As to the *B*-dependence of the MR (see Fig. 6(b) and (c)), a linear-in-*B* MR is also resolved when the field is anchored to the maximum angle ($\theta$ = 45°). With increasing *T*, the quadratic MR at low *B* is gradually restored while the high-*B* MR remains linear.

Very similar 'butterfly' AMR has been reported in the NLSM ZrSiS in which the Dirac nodes emerge in both the bulk and surface electronic states [11], in analogy with the Dirac-node-arc compound $PtSn_4$. In ZrSiS, the butterfly-like AMR was disassembled into two-fold and four-fold components whose weights both varied with *T* and *B* [11]. The butterfly-like AMR in ZrSiS persists up to *T* = 70 K though no butterfly-to-dumbbell change of AMR is observed.

III. Discussion

From this first study of the bulk transport and thermodynamic properties of single crystalline PdSn$_4$, a consistent picture for the electronic states near $\varepsilon_F$ has emerged. The $A$ coefficient of the $T^2$ resistivity, the electronic specific heat coefficient $\gamma$, and the slope of the TEP in the low-$T$ limit all show a significant enhancement relative to the values reported previously in the isostructural Dirac nodal arc semimetal PtSn$_4$. Such enhancement is typically attributed to heavier quasiparticle masses, a hypothesis seemingly confirmed by the observation of dHvA oscillation frequencies with corresponding $m^*$ values that are on average 3-4 times larger than those obtained in PtSn$_4$.

The origin of this mass enhancement is currently unknown. However, in nodal-line semimetals in general, it has been argued [19,20] that the specific properties of the band dispersions near $\varepsilon_F$, in particular the presence of the Dirac nodal line running parallel to the Fermi surface, can give rise to a Coulomb interaction, both onsite or nearest-neighbor, that remains partially unscreened. This residual Coulomb interaction can in turn lead to an *enhancement* in $m^*$ rather than a reduction (as seen, for example, at the isolated Dirac points in graphene), as well as to an increased tendency towards magnetic or charge density wave order, either in the bulk or on the surface [20,21]. The marked changes in TEP, $M(T)$, the AMR and $\rho_{xy}(B)$ around 25 K all suggest changes to sections of the Fermi surface, though clearly further work will be required to elucidate the origin of these features.

The strictly linear MR up to 55 T observed in PdSn$_4$ at certain field angles is another key finding of this work, and contrasts markedly with the quadratic MR seen in its sister PtSn$_4$. A natural question arises concerning the origin of this large LMR. Indeed, the weakly diamagnetic response in non-magnetic PdSn$_4$ would appear to rule out quantum electron-electron ($e$-$e$) interference effects, as proposed for the origin of the LMR in certain (geometrically constrained) ferromagnets [31,32].

Density fluctuations have also been proposed as the origin of a linear MR, this time in ultrahigh mobility GaAs quantum wells [33]. Such density fluctuations cause a gradient of the Hall voltage (which is linear in $B$) that will dominate the MR response and give rise to a LMR. However, this situation is unlikely to be applicable here because of the non-linear Hall resistivity $\rho_{xy}$ at low temperatures in this high-quality PdSn$_4$.

Another possible interpretation of LMR is the breakdown of weak-field magnetotransport near a quantum critical point [34]. At the critical point for a density-wave transition, for example, the weak-field regime of magnetotransport can collapse to zero with the size of the gap [34]. Indeed, in certain quantum materials exhibiting charge density wave (CDW) or spin density wave (SDW), e.g. LaAgSb$_2$ [27], Ca$_3$Ru$_2$O$_7$ [34] and the iron-based superconductors BaFe$_2$As$_2$ [35] and S-doped FeSe [36], LMR disappears at or above the CDW or SDW onset temperatures.

In TIs or TSMs, a LMR can result from a linear energy dispersion of the Dirac fermions [5,6,37,38]. There, the energy splitting between the lowest and the first Landau level (LL) can be described by $\Delta_{LL}$ = $v_F$(2e$\hbar$ B)$^{0.5}$. In the quantum limit, $\Delta_{LL}$ is much larger than both $\varepsilon_F$ and $k_BT$ and as such, all carriers are degenerate in the lowest LL. Thus, once above a certain critical field $B_C$, quantum transport will dominate and a non-saturating LMR is expected. In this case, a crossover from the low-$B$ quadratic dependence to the high-$B$ linear MR will occur at a specific field strength $B_C$ which follows the $T$-dependent relation [27]:

$$B_C = \frac{1}{2e\hbar v_F^2}(\varepsilon_F + k_BT)^2.$$

While neither the low-$B$ quadratic MR nor the above $T$-dependence of $B_C$ are seen experimentally, the butterfly AMR observed here for PdSn$_4$ has also been seen in the NLSM ZrSiS, and a LMR without quadratic $B$ dependence at low fields has been reported in other TSMs in which Dirac states were firmly established. We therefore tentatively attribute the LMR seen here to the existence of a Dirac-like linear dispersion in PdSn$_4$ too.

The change in the AMR symmetry with varying $T$ points to changes in the relative mobilities of different sections of the Fermi surface. The similarity in the dHvA frequencies found for both systems implies that the main features of the Fermi surface remain intact upon Pd substitution. It is nonetheless intriguing that the effective masses are significantly enhanced in PdSn$_4$, by a factor of 3-4 compared to PtSn$_4$. Enhancement by such a large factor would appear to rule out electron-phonon coupling as the responsible interaction and instead imply the presence of enhanced electron-electron correlations. These interesting many-body interactions, which have been ignored in most treatments of topological physics to date, certainly deserve more investigation, both theoretically and experimentally. Overall, our collective results suggest that PdSn$_4$ may be a new candidate topological material and a promising platform to study both Dirac physics and electron correlations in a single system.

IV. Materials and Methods

Single crystals of PdSn$_4$ were fabricated via self-flux methods. High-purity starting materials Pd (3N, 100 mosh from Alfa Aesar) and Sn (3N, from Aladdin) were mixed thoroughly with a molar ratio of 1:20. This process was handled in a glove box filled with protective argon gas (H$_2$O and O$_2$ contents below 0.1 ppm). Then the mixed materials were sealed in vacuumed quartz tubes (diameter ɸ= 2 cm). The tubes were heated to 673 Kelvin for 3 hours in high temperature box furnace and kept at this temperature for more than 7 days. After furnace cooling of the tubes to room temperature, bulk ingots were harvested. The redundant tin flux on the surface of crystals was dissolved in dilute hydrochloric acid. The actual stoichiometric ratio between Pd and Sn has been verified by energy dispersive X-ray spectroscopy (EDS) (see Supplementary Information).

Single crystal X-ray diffraction (XRD) measurements were performed at room temperature using a Rigaku diffractometer with Cu $K\alpha$ radiation and a graphite monochromator. A typical XRD pattern is shown in SI Fig. S1. Only the (0$\underline{2l}$0) peaks were observable, suggesting that the crystallographic *b*-axis is perfectly perpendicular to the surface facet of the crystal. All (0$\underline{2l}$0) peaks can be well indexed with the orthorhombic (Ccca, No. 68) space group, the same structure as PtSn$_4$. The crystal lattice parameter *b* is estimated as 11.451 $\pm$ 0.003 A, slightly larger than that in PtSn$_4$. Specific heat data were obtained using a relaxation method on a Quantum Design (QD) physical property measurement system (PPMS). Resistivity, magneto-resistivity and Hall data were measured by a standard four-probe method under various field/current configurations on the PPMS. High-field MR data up to 55 T were obtained from the pulsed magnetic field laboratory in Wuhan. The Hall measurements were conducted by an orthometric electric-contact configuration and data anti-symmetrization was performed based on reversing the field polarities. For thermoelectric power (TEP) measurements, a chip heater was attached to one end of the sample to generate a temperature gradient which was monitored by a constantan-chromel differential thermocouple, and the thermopower voltage was read out by a nanovoltmeter K2182A from Keithley Instruments. The magnetization data were measured using a QD Superconducting Quantum Interference Device (SQUID) magnetometer with fields up to 7 T. High-pressure electrical measurements (data shown in the SI) were carried out on PPMS using a piston cylinder type pressure cell (type: HPC-33, transmission medium: Daphne 7373) with a pressure up to 2.5 GPa.

**Acknowledgements**

The authors would like to thank C. M. J. Andrew for the critical reading of the manuscript and J. J. Feng for technical support. This work is sponsored by the National Key Basic Research Program of China (Grant No. 2014CB648400), and by National Natural Science Foundation of China (Grant No. 11474080, 11574097, 11611140101). Work in high field magnet laboratory was supported by the National Key Research and Development Program of China (Grant No.2016YFA0401704). X. X. would also like to acknowledge the financial support from an open program from Wuhan National High Magnetic Field Center (2015KF15).

**Author contributions**

C.Q.X., W.Z. grew the crystals and performed the materials composition and structural analyses. C.Q.X., W.Z., R. S., X. Z. X., Z. X. S., A.F.B., Z. D. H. performed and analysed the low field measurements. C.Q.X., J.H.W, J.L.Z. and Z.W.Z performed the high field measurements. X.X. and B.Q. conceived the project and analysed the results. C. Q. X. and W. Z. contributed equally to this work. N.E.H and X.X. wrote the manuscript with input from all authors.

Correspondence should be addressed to: X. F. Xu (xiaofeng.xu@cslg.edu.cn) or B. Qian (njqb@cslg.edu.cn)

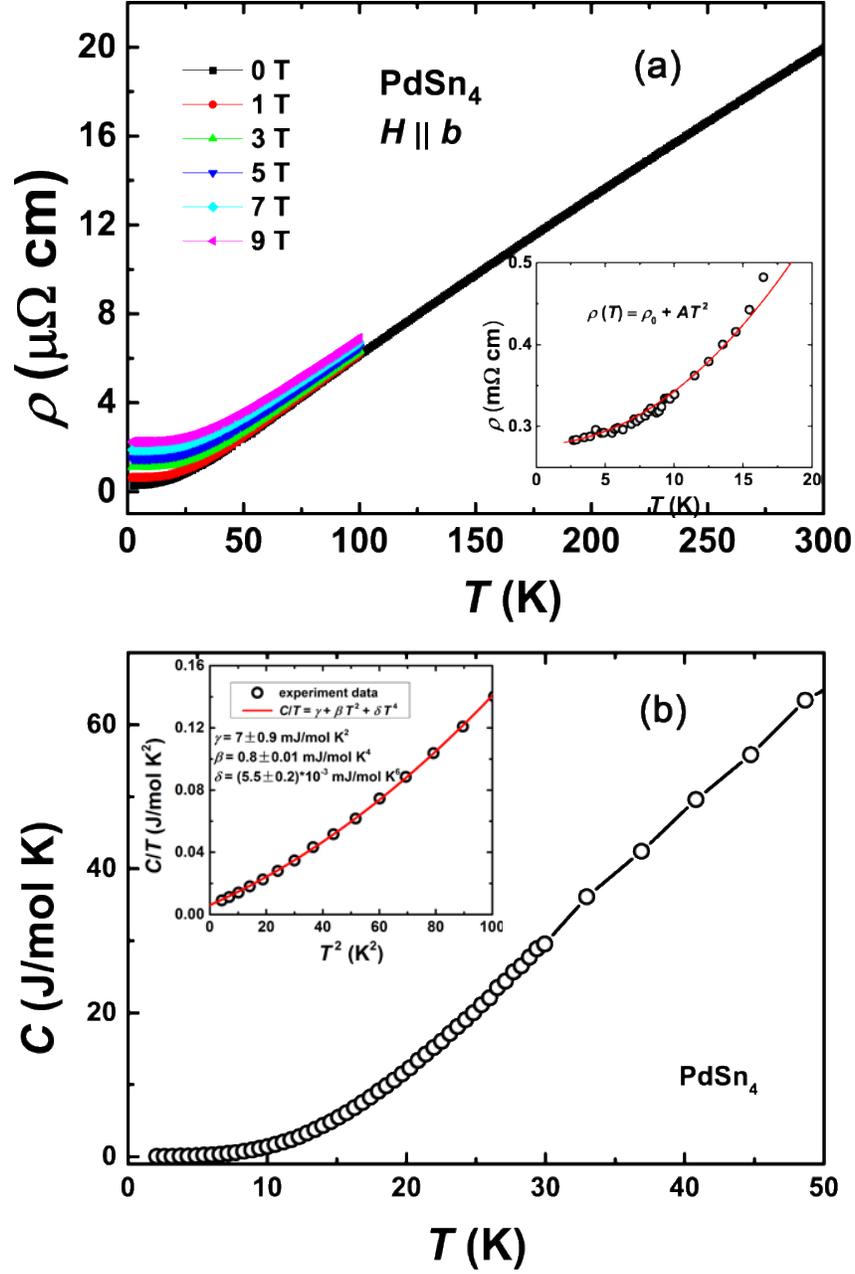

FIG. 1. (a) Temperature dependence of resistivity $\rho$ under different magnetic fields (H // b). Inset is an enlarged view of $\rho$ (T) in low temperature regime and the fit to $\rho(T) = \rho_0 + AT^2$. (b) Temperature dependence of specific heat C. Inset shows a fit to $C(T) = \gamma T + \beta_3 T^3 + \beta_5 T^5$.

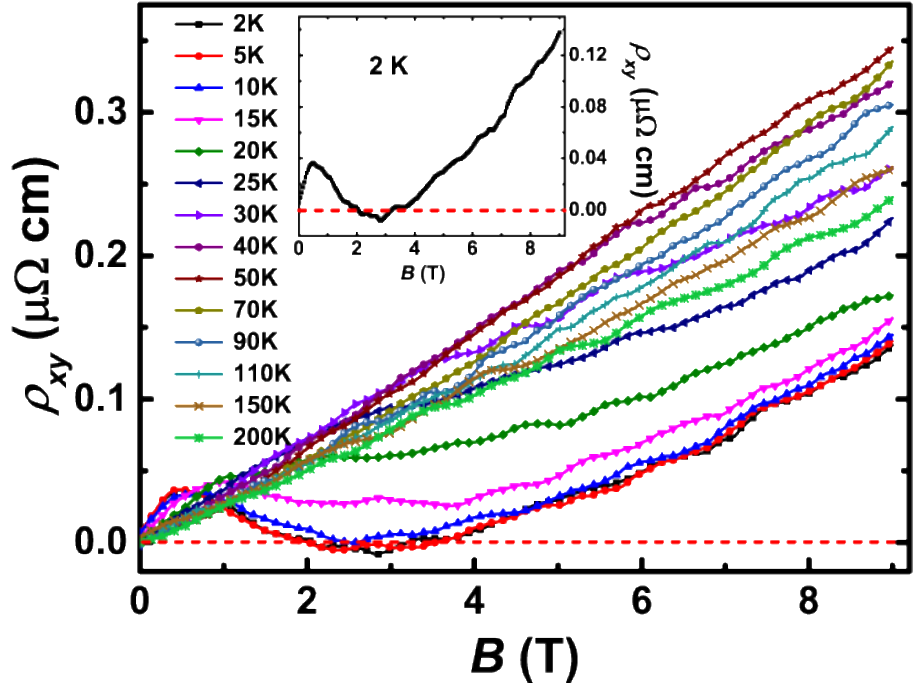

FIG. 2. *B*-dependence of the Hall resistivity $\rho_{xy}$ at different temperatures measured. Inset is an individual plot of $\rho_{xy}(B)$ at *T* = 2 K to highlight the presence of the two inflexion points. The horizontal dashed lines are guides to the eye and demonstrate the negative values in some field interval.

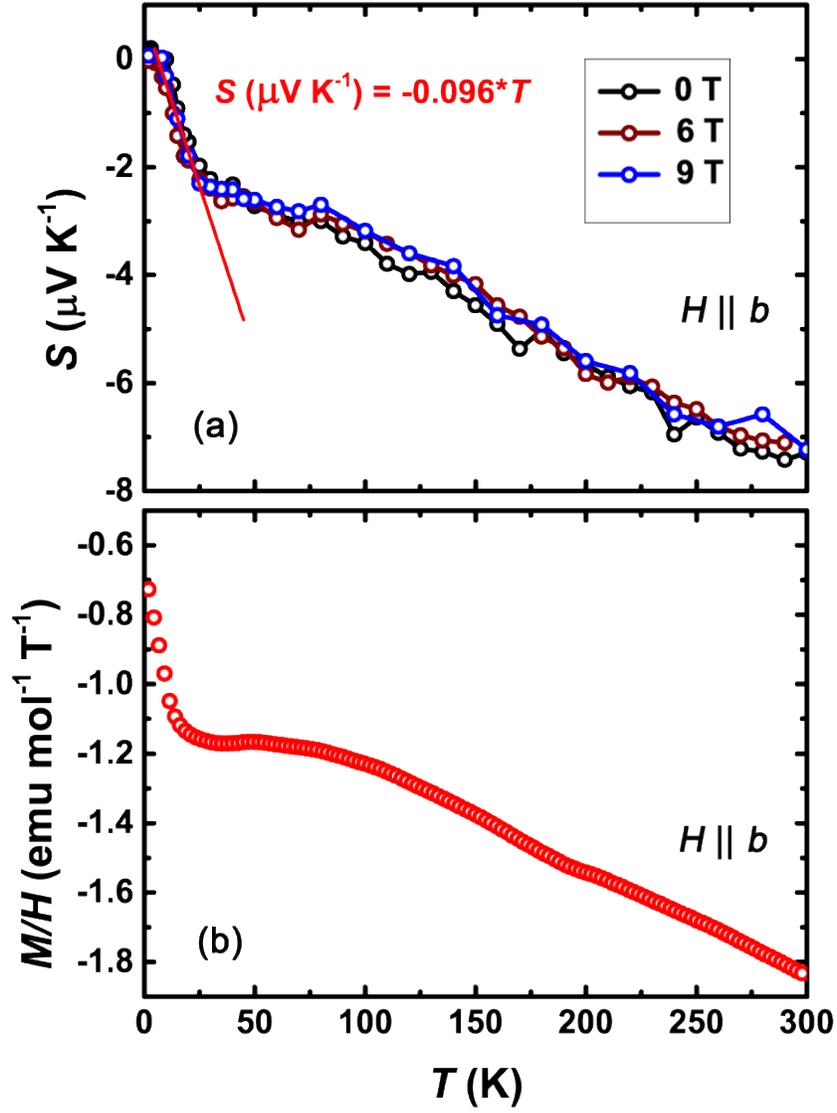

FIG. 3. (a) Temperature dependence of the TEP ($S$) under magnetic fields of 0, 6 T, 9T (H // $b$). The red solid line is a fit to the Sommerfeld theory of TEP for a single-band metal. (b) Temperature dependence of the magnetic susceptibility ($\chi$ = M/H) measured under a field of 0.5 T. $\chi(T)$ measured under other field strengths exhibit the same $T$-dependence (data not shown).

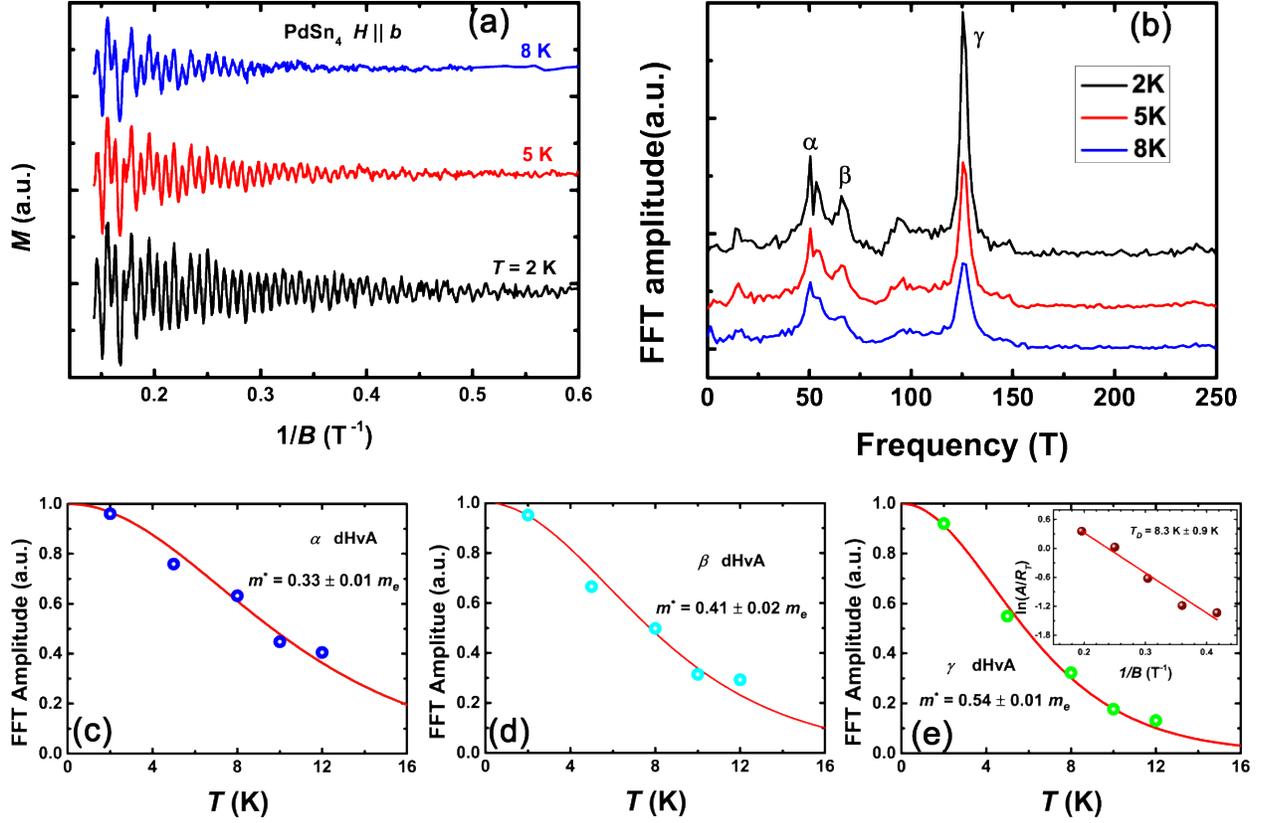

FIG. 4. (a) dHvA oscillations in magnetization at different temperatures (as labelled) after subtracting a polynomial background. The magnetic field is applied along the *b*-axis. (b) FFT spectra of the dHvA data at the same temperatures. At least three extremal Fermi surface cross-sectional areas are identified by the peaks in the FFT spectra, i.e. $\alpha$ (50 T), $\beta$ (66 T), $\gamma$ (125 T). (c-e) Temperature dependence of the normalized FFT amplitude which is proportional to the thermal damping factor $R_T$. The red solid lines are fits to the Lifshitz-Kosevich formula, which is used to extract the cyclotron effective mass $m^*$. Inset in (e) shows a typical Dingle plot used to obtain the Dingle temperature $T_D$. The error bars for $m^*$ and $T_D$ are obtained by the standard deviations during the above fits.

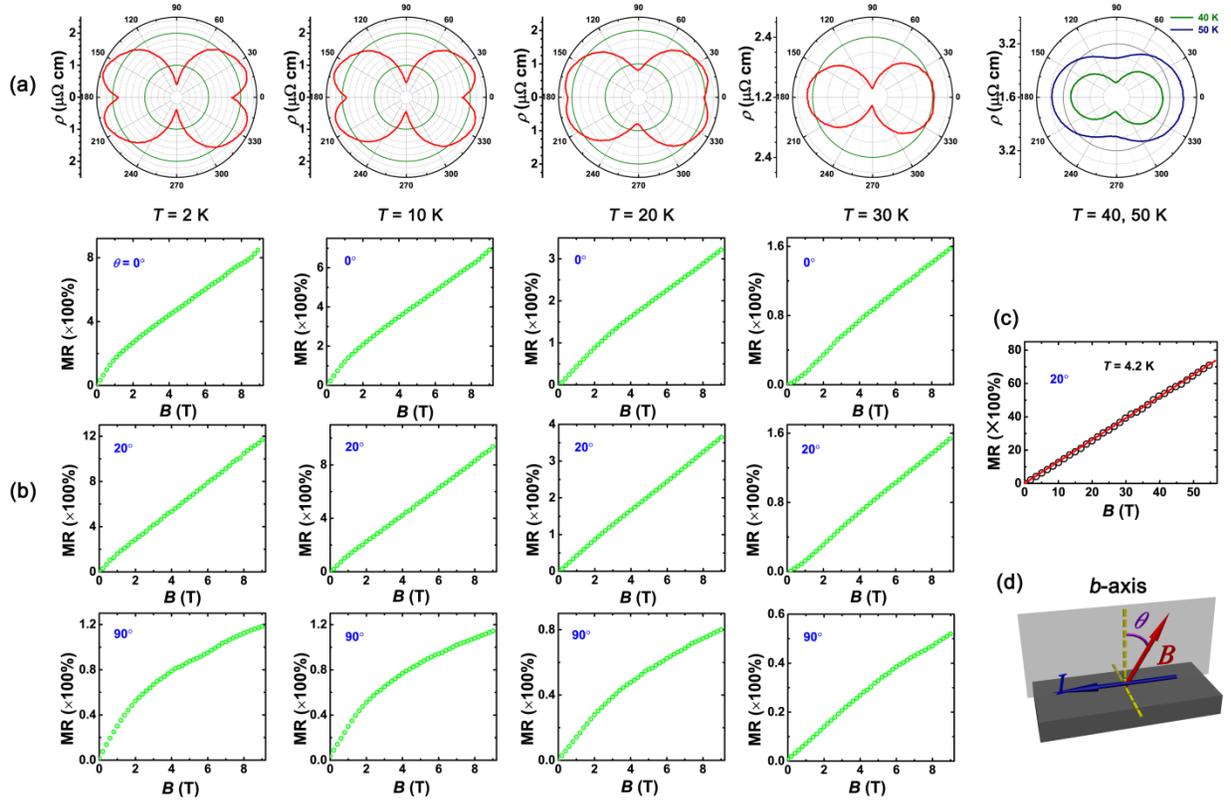

FIG. 5. (a) Angle dependence of resistivity $\rho$ under $B = 9$ T for several selected temperatures ($B // bI$ plane, see the field rotation configuration in (d)). (b) Field dependence of magnetoresistance (MR =$(\rho(B)-\rho(0))/\rho(0)$) at some specific angles. (c) MR measured under a pulsed magnetic field up to 55 T at $\theta = 20°$. The red solid line is a guide for the linear $B$ dependence of MR. (d) Schematic drawing of the field/current configurations. Field is constrained to a plane formed by the current flow direction and the crystalline $b$-axis. $\theta = 0°$ corresponds to $B // b$-axis.

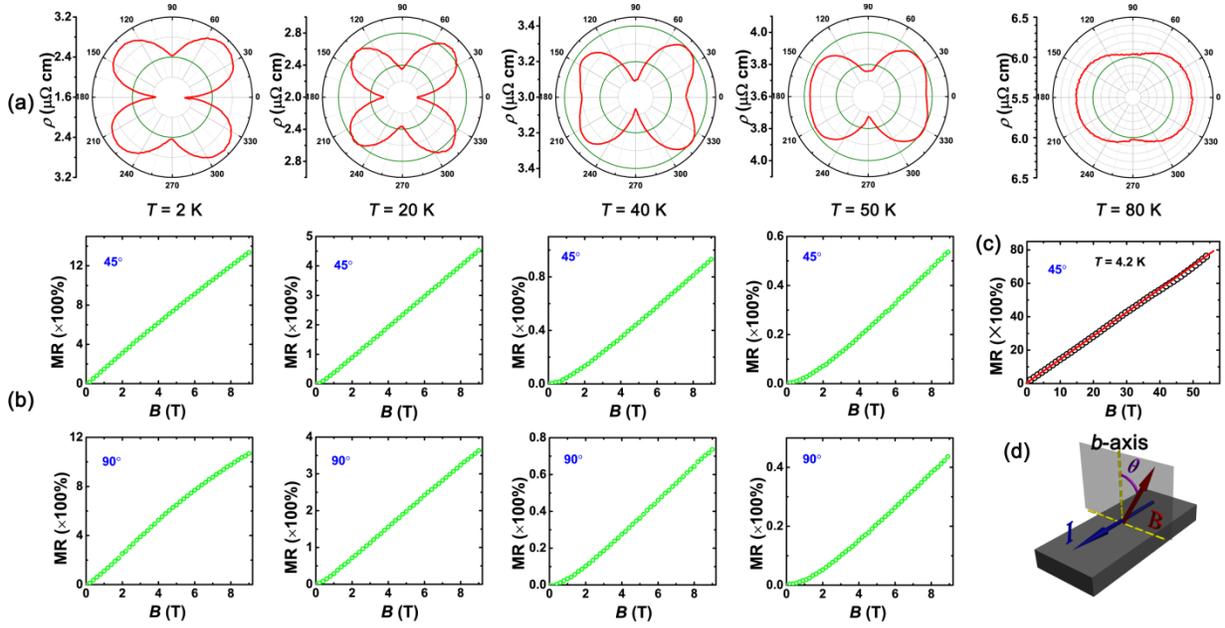

FIG. 6. (a) Angle dependence of resistivity $\rho$ under $B$ = 9 T for selected temperatures ($B \perp I$, see the field configuration in (d)). (b) Field dependence of magnetoresistance at angle $\theta$ = 45° and 90°. (c) MR measured under a pulsed magnetic field up to 55 T with $\theta$ = 45°. The red solid line is a guide for the linear $B$ dependence of MR. (d) Schematic view of the setup. Field rotates within the plane perpendicular to the current flow. $\theta$ = 0° corresponds to $B$ // $b$-axis.

TABLE I. Parameters obtained from dHvA oscillations of a PdSn$_4$ single crystal. $F$ is the peak frequency in the FFT spectra. $A_e$ is the extremal Fermi surface cross-sectional area calculated from the Onsager relation and $k_F$ is the corresponding Fermi wave vector. $m^*$ and $m_e$ are the effective electron mass and the bare electron mass, respectively. $T_D$ is the Dingle temperature. $v_F$ and $E_F$ are the Fermi velocity and Fermi energy, respectively. $\tau$ is the relaxation time, $\ell$ is the mean free path, and $\mu$ is the mobility, of charge carriers.

| dHvA | ($H // b$) | | PdSn$_4$ |
| --- | --- | --- | --- |
| Parameter | $\alpha$ | $\beta$ | $\gamma$ |
| $F$ (T) | 50 | 66 | 125 |
| $A_e$ (Å$^{-2}$) | $4.8 \times 10^{-3}$ | $6.3 \times 10^{-3}$ | $1.2 \times 10^{-2}$ |
| $k_F$ (Å$^{-1}$) | 0.039 | 0.045 | 0.062 |
| $m^*/m_e$ | $0.33 \pm 0.01$ | $0.41 \pm 0.02$ | $0.54 \pm 0.01$ |
| $T_D$ (K) | $10.7 \pm 0.5$ | $12.5 \pm 1.8$ | $8.3 \pm 0.9$ |
| $v_F$ (cm s$^{-1}$) | $(1.4 \pm 0.07) \times 10^7$ | $(1.3 \pm 0.1) \times 10^7$ | $(1.32 \pm 0.03) \times 10^7$ |
| $E_F$ (meV) | $35 \pm 2$ | $37 \pm 2$ | $54 \pm 2$ |
| $\tau$ (s) | $(1.1 \pm 0.1) \times 10^{-13}$ | $(9.7 \pm 1.7) \times 10^{-14}$ | $(1.5 \pm 0.2) \times 10^{-13}$ |
| $\ell$ (nm) | $16 \pm 2$ | $12 \pm 3$ | $19 \pm 3$ |
| $\mu$ (cm$^2$V$^{-1}$s$^{-1}$) | $610 \pm 50$ | $420 \pm 80$ | $480 \pm 50$ |